\documentclass{aastex}

\shorttitle{L-Band Photometry of Brown Dwarfs}
\shortauthors{Stephens et al.}

\begin{document}

\title{L-Band Photometry of L and T Dwarfs}

\author{Denise C. Stephens}
\affil{Astronomy Department, New Mexico State University, Box 30001, MSC 4500, Las Cruces, NM 88012}
\email{densteph@nmsu.edu}

\author{Mark S. Marley}
\affil{Space Sciences Division, NASA/Ames Research Center, Moffett Field, CA 94035, and}
\affil{Astronomy Department, New Mexico State University, Box 30001, MSC 4500, Las Cruces, NM 88012}
\email{mmarley@nmsu.edu}

\author{Keith S. Noll}
\affil{Space Telescope Science Institute, 3700 San Martin Drive, Baltimore, MD 21218}
\email{noll@stsci.edu}

\author{Nancy Chanover}
\affil{Astronomy Department, New Mexico State University, Box 30001, MSC 4500, Las Cruces, NM 88012}
\email{nchanove@nmsu.edu}

\begin{abstract}
We present {\it K}- and {\it L}-band photometry obtained with the Keck I telescope for a representative sample of L and T dwarfs.  These 
observations were motivated in part by the dominant role $\rm H_{2}O$ and $\rm CH_{4}$ play in shaping the flux near 2 and 3 $\mu$m and 
by the potential use of these bands as indicators of spectral class in the infrared.  In addition, these observations
aid the determination of the bolometric luminosity of L and T dwarfs.  Here we report the {\it K}, {\it L$^\prime$} and $\rm {\it L_s}$ magnitudes 
of our objects and the trends observed 
in the ({\it K-L$^\prime$}) and ({\it K}-$\rm {\it L_{s}}$) colors as a function of L- and T-dwarf spectral class.  We compare these colors with 
theoretical models, derive a relationship between effective temperature and L-spectral class, and compare our temperature estimates with others.  
\end{abstract}


\section{Introduction}

The discovery of hundreds of ultra low-mass dwarfs and substellar objects by the 2 Micron All Sky Survey (2MASS), the Deep Near
Infrared Southern Sky Survey (DENIS) and the Sloan Digital Sky Survey (SDSS) has provided a large data base from which to select
L and T dwarfs for further study.  The L dwarfs identified by these surveys have been classified using far optical ($\sim$ 0.6-1.0 $\rm \mu m$) 
spectroscopy and most have published {\it J} (1.1-1.4 $\rm \mu m$), {\it H} (1.5-1.8 $\rm \mu m$) and {\it K} band (2.0-2.4 $\rm \mu m$)
magnitudes \citep{del97, kir99, mar99, fan00, giz00, kir00}.  In addition several groups have used near-infrared spectroscopy from 1 to 2.5 
$\rm \mu m$ to classify and study L dwarfs \citep{del97, del99, tok99, mcl00, leg00, leg01, rei01, tes01}.  Comparatively, little work has been 
done beyond 2.5 $\rm \mu m$, where to date only six L dwarfs have published {\it L}-band magnitudes \citep{jon96, leg98, tok99, leg01}.  Since brown 
dwarfs emit most of their flux in the near infrared, observations beyond 2.5 $\rm \mu m$ are useful to the determination of the bolometric 
luminosity of these objects.  In addition water ($\rm H_{2}O$), carbon monoxide (CO), and methane 
($\rm CH_{4}$) which are among the most abundant atmospheric species after hydrogen (H$_2$) are important opacity sources at these 
wavelengths.

Indeed $\rm H_{2}O$ and $\rm CH_{4}$ opacities dominate the {\it L}-band (2.5-4.1 $\rm \mu m$)
radiative transfer.  In particular the $\rm {\it L_s}$ band (2.5-3.5 $\rm \mu m$) overlaps the strong $\rm \nu_{3}$ 
fundamental band of $\rm CH_{4}$ at 3.3 $\rm \mu m$, which is observable in some of the late L dwarfs and all of the T dwarfs \citep{nol00}.  
Because chemical equilibrium favors $\rm CH_{4}$ over CO with falling atmospheric temperature \citep{feg96}, the abundance of $\rm CH_{4}$
helps to constrain the effective temperature ($\rm T_{eff}$).  In the {\it L$^\prime$} band (3.5-4.1 $\rm \mu m$) $\rm H_{2}O$ alone is the dominant
opacity source.  Atmosphere models \citep{mar01} 
suggest that the ({\it K-L$^\prime$}) color of L dwarfs is relatively insensitive to the details of cloud models, or even the absence of clouds, 
making this color a potentially interesting diagnostic of $\rm T_{eff}$.  When used together, the $\rm {\it L_s}$ and {\it L$^\prime$} bands 
can probe for the appearance of $\rm CH_{4}$ in the latest L dwarfs, identify for trends in color which are a function of spectral type and temperature, 
and provide flux measurements for use in estimating the bolometric luminosity of these objects.  Here we report {\it L}-band observations for a 
sample of L and T dwarfs, their observed trends, and comment on the L-dwarf $\rm T_{eff}$ scale.

\section{Observational Technique}

The data were acquired using the Near Infrared Camera, NIRC \citep{mat94} on Keck I at the W.M. Keck Observatory located in Mauna Kea, Hawaii 
on April 13th and 14th, UT 2000.  Photometry was obtained using {\it K}, $\rm {\it K_s}$, $\rm {\it L_s}$, {\it L$^\prime$} and a 3.3 $\rm \mu m$
narrow band filter\footnote{The narrow band results will be presented elsewhere.} for 18 
L dwarfs and 5 T dwarfs which were observable at a low airmass, and chosen to span the L-dwarf spectral sequence.  Where multiple objects of a 
given spectral 
type were available to observe, preference was given to those objects which were most typical in their ({\it J-K}) and ({\it H-K}) colors respective 
to spectral type \citep{kir00}.  The resulting data set includes objects with spectral types ranging from L0 to L8 in the \citet{kir99} classification as
well as five T dwarfs ranging from T0 to T6 in the A. Burgasser (2001, private communication) classification.
Table 1 lists the targets, their coordinates, observed magnitudes and colors.  Both nights were photometric with
seeing ranging between 0.35\arcsec and 0.6\arcsec.  Since the radiation background from the sky is considerably larger at 3 $\rm \mu m$ and beyond,
integration times were limited to 0.012 and 0.03 seconds in the {\it L$^\prime$} and $\rm {\it L_s}$ bands.  To achieve these integration times, a 
64 $\times$ 64 subsection of the chip was read out using the pat4xa timing pattern.  Standard chopping and nodding procedures were also employed 
to minimize errors resulting from temporal variations in the sky radiation background.  

To place our photometry onto a standard infrared photometric system, standard stars were chosen from the UKIRT Fundamental and Extended List 
\citep{haw01} and were observed at the same airmass directly before or after the target objects.  In the case where a UKIRT standard of the same
airmass was not available to observe, standard stars were then chosen from the \citet{per98} NICMOS standard list and their magnitudes were
converted to the UKIRT system using the formula given by \citet{haw01}.   These lists provided standard {\it K}-band magnitudes, but not {\it L$^\prime$}
or $\rm {\it L_s}$ magnitudes.  To determine a standard magnitude for our {\it L}-band observations we observed only main sequence standard stars 
with spectral types between A0V and G6V, and estimated a standard {\it L$^\prime$} or $\rm {\it L_s}$ magnitude for each star
using the intrinsic colors given by \citet{bes88} for a main sequence star of that spectral type.  Since the ({\it K-L$^\prime$}) and ({\it K-L}) values of 
these stars vary
between 0.0 for A0V to 0.05 for G6V, we estimate that this technique produces no more than a 5\% error in our magnitude 
calculation and this  error has been included in the final {\it L$^\prime$} and $\rm {\it L_s}$ magnitude tabulation (Table 1).  
As an additional check standard stars that deviated largely from the mean {\it L$^\prime$} and $\rm {\it L_s}$
zeropoint correction determined from all of the standard stars were discarded.  This resulted in the elimination of no more than one standard star each 
night and
we are confident that this technique produces valid {\it L$^\prime$} and $\rm {\it L_s}$ magnitudes for our objects.  

The standard stars used to calibrate the photometry in Table 1 are on the UKIRT IRCAM3 natural system.
Convolutions of the filter transmission profiles for the NIRC and IRCAM3 filters with model spectra reveals a constant zeropoint
offset between the two systems with no dependence on spectral type.  Thus the photometry presented in this paper is on 
the UKIRT IRCAM3 system.  It is important to note 
that this is not the same photometric system as the Mauna Kea Observatory near-infrared system (MKO-NIR).  While the {\it K} band magnitudes
for our L dwarfs can be placed on the MKO-NIR system using the transformation equations provided by \citep{haw01}, there is not a conversion
equation for the {\it L$^\prime$} magnitudes.  A comparison of the  {\it L$^\prime$} filters for both the NIRC and MKO-NIR systems reveals that the 
MKO-NIR  filter extends farther into the blue from 3.4 to 4.1 $\rm \mu m$ while the NIRC filter only extends from 3.5 to 4.1 $\rm \mu m$.  This 
results in the MKO-NIR filter being more sensitive to the appearance of $\rm CH_{4}$ absorption at 3.3 $\rm \mu m$.  The net effect is that for 
the hottest
L dwarfs, our {\it L$^\prime$} magnitudes are almost identical to the MKO-NIR magnitudes.  For the latest L dwarfs where $\rm CH_{4}$ is just 
beginning to affect the spectrum, our {\it L$^\prime$} magnitudes are 10\% brighter, and for the T dwarfs where $\rm CH_{4}$ absorption is strongly
established, our {\it L$^\prime$} magnitudes are 20\% brighter (differences of 0.1 and 0.2 mag respectively).  Thus any comparison of the 
{\it L$^\prime$} magnitudes from these two systems needs to take this offset into account.

As a final comment on the magnitudes presented here, we need to mention that although observations were acquired using both the 
{\it K} and $\rm {\it K_s}$ filters,
we found that in general the {\it K}-band magnitude differed from the $\rm {\it K_s}$-band magnitude by no more than 0.05 for the 
L dwarfs and 0.1 for the T dwarfs.  In addition since transformation equations from the UKIRT IRCAM3 system to the other standard infrared systems
use the UKIRT {\it K}-band magnitudes for conversion and not $\rm {\it K_s}$, we present only the {\it K}-band magnitudes.

\section{Infrared Colors}

Figure 1 presents the ({\it K-L$^\prime$}) color of our objects as a function of L- and T-dwarf spectral class.
The closed squares represent the objects from Table 1, and the open circles represent ({\it K-L$^\prime$}) colors for six L dwarfs
by \citet{leg01}.  In general the ({\it K-L$^\prime$}) color gradually increases with later L- and T-dwarf spectral type.  A linear least 
squares fit of the data to the L dwarf index
shows that the ({\it K-L$^\prime$}) colors are reasonably well fit (correlation coefficient, $r = 0.80$) by a line with a slope 
of 0.092 and an intercept point of 0.87.  For L dwarfs the increase in ({\it K-L$^\prime$}) color is due to increasing molecular absorption in the 
{\it K} band.  As the $\rm T_{eff}$ decreases
with later spectral type, the wings of the $\rm H_{2}O$ band at 1.9 $\rm \mu m$ and pressure induced $\rm H_{2}$-$\rm H_{2}$ absorption 
together suppress the {\it K}-band flux.  In the {\it L$^\prime$} band the principle opacity source is $\rm H_{2}O$ alone.
As the temperature decreases, the mean opacity increases more slowly in the {\it L$^\prime$} band than in the {\it K} band.
The net result is that the total flux emitted in the {\it K} band falls faster than the total flux emitted in the {\it L$^\prime$} band
with decreasing temperature.  This produces the redder ({\it K-L$^\prime$}) colors with later L-dwarf spectral type.

A similar mechanism is responsible for the gradual reddening of the T dwarfs.  For these objects $\rm CH_{4}$ joins $\rm H_{2}O$ and 
$\rm H_{2}$-$\rm H_{2}$ as the principle opacity sources.
By definition, the transition from the L to T dwarfs occurs when the $\rm \nu_{2}+\nu_{3}$ combination band of $\rm CH_{4}$ first appears at 
2.2 $\rm \mu m$ \citep{kir99}.  A linear least squares fit ($r = 0.97$) of the T dwarf data produces a slightly steeper line with a slope of 0.161 and 
an intercept point of -0.19.  The change in slope from the L to T dwarfs suggests that the additional appearance of $\rm CH_{4}$ opacity
further suppresses the flux in the {\it K} band, resulting in a steeper ({\it K-L$^\prime$}) color increase throughout the T dwarfs.  Since the 
abundance of $\rm CH_{4}$ varies with $\rm T_{eff}$ and it is the dominant opacity source for T dwarfs at {\it K} band, but not
{\it L$^\prime$}, the ({\it K-L$^\prime$}) color should prove to be a good diagnostic for estimating the $\rm T_{eff}$ of T dwarfs.

Figure 2 presents the ({\it K}-$\rm {\it L_s}$) color of our objects as a function of L- and T-dwarf spectral class.
Like the ({\it K-L$^\prime$}) color, the increase in ({\it K}-$\rm {\it L_s}$) through the L-dwarf spectral sequence is due to increasing
$\rm H_{2}O$ and $\rm H_{2}$-$\rm H_{2}$ absorption in the {\it K} band.  A linear least squares fit applied to this data ($r = 0.50$) produces
a line with a slope of 0.06 and an intercept point of 0.14.  It is interesting to note that there appears to be substantially more scatter in the
({\it K}-$\rm {\it L_s}$) color for the later L dwarfs (and early T dwarfs) as compared to the early L dwarfs.  This scatter may arise from differences 
in cloud cover \citep{gel01}, or possibly atmospheric dynamics or metallicity producing greater variability in $\rm CH_{4}$ 
abundance in these objects.  

For the T dwarfs $\rm CH_{4}$ absorption definitely occurs both at 3.3 $\rm \mu m$ and 2.2 $\rm \mu m$ which results in a suppression of the 
flux in the $\rm {\it L_s}$ and {\it K} bands.  Since the $\rm \nu_{3}$ fundamental band of $\rm CH_{4}$ at 3.3 $\rm \mu m$ is a factor of 100 times
stronger than the $\rm \nu_{2}+\nu_{3}$ combination band at 2.2 $\rm \mu m$, there is a greater suppression of the flux in the $\rm {\it L_s}$ band 
than in the {\it K} band which results in a bluer color for the T dwarfs.  If $\rm CH_{4}$ absorption were to gradually appear in the latest 
L dwarfs with falling $\rm T_{eff}$, we would expect the ({\it K}-$\rm {\it L_s}$) color of the late L dwarfs to gradually transition into the bluer color 
of the early T dwarfs.  However, the abrupt transition in ({\it K}-$\rm {\it L_s}$) from the latest L dwarfs to the earliest T dwarfs suggests that
perhaps very early T dwarfs, or very late L dwarfs have yet to be discovered, or that substantial $\rm {\it L_s}$ band $\rm CH_{4}$ absorption 
appears rapidly over a very small $\rm T_{eff}$ range.

Note that the two L1 dwarfs both have very blue ({\it K}-$\rm {\it L_s}$) colors, but unexceptional ({\it K-L$^\prime$}) colors.  These 
two objects were also unusually faint in narrow band 3.3 $\rm \mu m$ (NIRC pahl filter) images obtained the same night.
This suggests that another opacity source is affecting the flux in the $\rm {\it L_s}$ band for these two
objects.    Clearly 3 to 4 $\rm \mu m$ spectroscopy is needed to constrain what is suppressing the flux in the 
$\rm {\it L_s}$ band.  We are confident that the colors of the L1 dwarfs are not due to photometric errors since these objects were observed
at different times during the night and with different standard stars.  One of the L1 dwarfs, DENIS J1441-0945, is a binary system and if the two
components of the system have different $\rm {\it L_s}$ magnitudes this may be the cause of the color discrepancy.  Indeed it does appear
that the pair is of unequal brightness although they are too close for individual photometry\footnote{Magnitudes for this object in Table 1
are for the combined pair.}.  However this would imply that
the other L1 dwarf is also a binary system and we see no evidence of multiplicity in our 0.5\arcsec  images of 2MASSW J1035+2507.

Finally Figure 3 presents a ({\it K}-$\rm {\it L_s}$) verses ({\it K-L$^\prime$}) color-color diagram with the data from Table 1.  The progressive
reddening in ({\it K-L$^\prime$}) is apparent as is the greater variability in ({\it K}-$\rm {\it L_s}$) at the L to T transition.
Note that the L7 object, 2MASSI J1526+2043, has the same ({\it K-L$^\prime$}) and ({\it K}-$\rm {\it L_s}$) colors as an L4 dwarf.  This L7 also has 2MASS 
({\it J}-$\rm {\it K_s}$), ({\it J-H}) and ({\it H} -$\rm {\it K_s}$) colors which are most similar to an L3 or L4 dwarf \citep{kir00}.  It is 
possible that this L7 dwarf has been misclassified and this should be kept in mind when studying its location in Figures 1, 2 and 3.

\section{Comparison to Models}

A complete analysis of the L-dwarf $\rm T_{eff}$ scale requires consideration of infrared and optical spectra as well 
as photometry.  However, we can use the results presented here to draw some tentative conclusions by comparing to 
atmospheric models of \citet{mar01}.  These models account for sedimentation of atmospheric condensates as described in \citet{ack01}.
\citet{mar01} found that the models with the precipitation efficiency 
factor $f_{\rm rain} =5$ best fit the 
({\it J-K}) colors of the L dwarfs and we use those models here.
The models suggest that the ({\it K-L$^\prime$}) color of dusty L-dwarf models is
insensitive to details of cloud structure or
gravity.  For 1300 K $\rm \la T_{eff} \la$ 2100 K
a remarkably accurate linear expression ($r = 0.99$) reproduces the ({\it K-L$^\prime$})
model colors as a function of $\rm T_{eff}$.  
Equating this expression with the regression line derived in Figure 1 for the ({\it K-L$^\prime$}) data
gives the following estimate of L dwarf $\rm T_{eff}$
as a function of \citet{kir99} spectral type $\rm L_{K}$:
$$\rm T_{eff} = 2220 - 100*L_K  \eqno(1)$$
where $\rm T_{eff}$ is the effective temperature in Kelvin, and
$\rm L_{K}$ = 0 to 8.  This equation was
fit using only models appropriate for $M\approx 30\,\rm M_J$,
however the colors depend only slightly on mass for this temperature range, and introduce 
only a few hundreths variation in the color difference at a fixed $\rm T_{eff}$. 

This equation is presented for its heuristic value as an indicator of the $\rm T_{eff}$ range of L-dwarfs
and not as a definitive analysis.  Nevertheless it is interesting to make some
comparisons.  The earliest L dwarfs appear to have $\rm T_{eff}\sim 2100$ to $2200\,\rm K$ and the L-dwarf regime extends down
to about $\rm T_{eff}\sim 1400\,\rm K$.  On average each \citet{kir99} spectral type spans about 100 K in $\rm T_{eff}$.

The $\rm T_{eff}$ derived here for L dwarfs can be compared to the results predicted by other research groups.
\citet{leg01} compares low resolution spectra and colors of a selection of L dwarfs to models of 
Allard, Hauschildt, \& Schweitzer (2000) to derive $\rm T_{eff}$ for several objects.  
Their spectral synthesis fitting gave an estimated $\rm T_{eff}$  of 1800-1900 K
for the entire L dwarf sequence (L0 to L8).  Comparison of observed fluxes to evolution models by \citet{cha00}
resulted in $\rm T_{eff}$ ranging from 2050 - 2350 K for L0 to 1400 to 1600 K for L7.
\citet{bas00} proposed that $\rm T_{eff}$ on the \citet{mar99} scale varies from about 2200 K (L0) to 1600 K (L6).
The above $\rm T_{eff}$ estimates are all
based on fits to the ``cleared dust" models of \citet{all00} which do not include dust in the radiative transfer.
Since the \citet{mar01} models do include dust opacity, accounting for sedimentation, the atmosphere models are
warmer at fixed $\rm T_{eff}$ than comparable cleared-dust models by several hundred degrees.  Thus the $\rm T_{eff}$ range implied by 
Eq. 1 is cooler and may be more reasonable.  

Our estimated effective temperatures are somewhat higher, particularly for the later-type objects, then those of \citet{pav00} who derive temperatures
of 2200 K for M9.5 to 1200 K for L7 based on spectral synthesis fitting of far optical spectra using dusty atmosphere models.  We note that their 
fitted $\rm T_{eff}$ depends sensitively upon the treatment of dust and alkali gas opacity as well as the inclusion of an unknown ``additional opacity''.  
In contrast the models of \citet{mar01} fit both the optical and near-infrared colors of L dwarfs by including a more physically-based description of
condensate sedimentation and properly accounting for chemical equilibrium with cloud formation.  Both dust sedimentation and the chemical equilibrium 
treatment of alkali metal opacity become more important with later L spectral type \citep{mar01}.  These effects are likely responsible for the 
discrepancies between our results.

\section{Conclusions}

We find that the ({\it K-L$^\prime$}) and ({\it K}-$\rm {\it L_s}$) colors become redder with later L-dwarf spectral type.
Similarly the ({\it K-L$^\prime$}) color of the T dwarfs steadily reddens by a full magnitude over the range of objects observed, and this color
may prove useful in the future classification and determination of effective temperature in T dwarfs.  We find that ({\it K}-$\rm {\it L_s}$) for the 
early T dwarfs is bluer than ({\it K}-$\rm {\it L_s}$) for the latest L dwarfs due to $\rm CH_{4}$ absorption but that a gradual transition between 
the two spectral classes is not seen.  This suggests that perhaps very early
T dwarfs or very late L dwarfs have yet to be found or that $\rm CH_{4}$ absorption appears rapidly over a very small effective temperature range.  
By using the observed ({\it K-L$^\prime$}) colors and theoretical models which include dust opacity and account for 
cloud sedimentation \citep{mar01}, we estimate an $\rm T_{eff}$ range of 2200 K to 1400 K for the entire L-dwarf spectral sequence (L0 to L8).

\acknowledgments

We thank Sandy Leggett for helpful comments and advice.  D.S. and M.M. acknowledge support from NASA grants 
NAG2-6007 and NAG5-8919 and NSF grants AST-9624878 and AST-0086288.
The data presented herein were obtained at the W.M. Keck Observatory, which is operated as
a scientific partnership among the California Institute of Technology, the University of
California, and the National Aeronautics and Space Administration.  The Observatory was
made possible by the generous financial support of the W.M. Keck Foundation.
We would also like to thank instrument specialists Bob Goodrich and Randy Campbell for 
their support and assistance in using the NIRC instrument.

\clearpage

\begin{deluxetable}{ccrrrrr}

\tabletypesize{\scriptsize}
\tablecaption{Colors and Magnitudes of Target Objects \label{tbl-1}}
\tablewidth{0pt}
\tablehead{
\colhead{Object Name} & \colhead{Spec\tablenotemark{a}}   & \colhead{K mag}   &
\colhead{L$^\prime$ mag} & 
\colhead{$\rm L_s$ mag}  & \colhead{K$-$L$^\prime$} & \colhead{$\rm K-L_s$}}
\startdata
TVLM 513$-$46546 &M8.5 &10.74 $\pm$ 0.01 &9.92 $\pm$ 0.05 &10.57 $\pm$ 0.05 &0.82 $\pm$ 0.06 &0.17 $\pm$ 0.05 \\
DENIS J1159384$+$005727 &L0 &12.80 $\pm$ 0.01 &11.87 $\pm$ 0.05 &12.60 $\pm$ 0.04 &0.93 $\pm$ 0.05 &0.20 $\pm$ 0.04 \\
2MASSW J1035245$+$250745 &L1 &13.33 $\pm$ 0.02 &12.56 $\pm$0.05 &14.46 $\pm$ 0.06 &0.78 $\pm$ 0.05 &$-$1.12 $\pm$ 0.07 \\
DENIS J1441373$-$094559\tablenotemark{b} &L1 &12.66 $\pm$ 0.03 &11.78 $\pm$ 0.06 &13.74 $\pm$ 0.06 &0.88 $\pm$ 0.07 &$-$1.08 $\pm$ 0.07 \\
2MASSW J1411175$+$393636 &L1.5 &13.25 $\pm$ 0.01 &12.22 $\pm$ 0.04 &13.08 $\pm$ 0.03 &1.03 $\pm$ 0.04 &0.17 $\pm$ 0.03 \\
2MASSW J0928397$-$160312 &L2 &13.63 $\pm$ 0.03 &12.72 $\pm$ 0.08 &13.38 $\pm$ 0.06 &0.91 $\pm$ 0.09 &0.25 $\pm$ 0.07 \\
2MASSW J1338261$+$414034 &L2.5 &12.73 $\pm$ 0.02 &11.72 $\pm$ 0.05 &12.49 $\pm$ 0.02 &1.01 $\pm$ 0.05 &0.24 $\pm$ 0.03 \\
2MASSW J1615441$+$355900 &L3 &12.92 $\pm$ 0.02 &11.61 $\pm$ 0.05 &12.60 $\pm$ 0.05 &1.31 $\pm$ 0.06 &0.32 $\pm$ 0.06 \\
2MASSW J1155009$+$230706 &L4 &14.07 $\pm$ 0.01 &12.75 $\pm$ 0.05 &13.70 $\pm$ 0.05 &1.33 $\pm$ 0.05 &0.38 $\pm$ 0.05 \\
2MASSW J1246467$+$402715 &L4 &13.22 $\pm$ 0.02 &11.88 $\pm$ 0.04 &12.79 $\pm$ 0.03 &1.33 $\pm$ 0.04 &0.42 $\pm$ 0.03 \\
2MASSW J1112257$+$354813 &L4.5 &12.69 $\pm$ 0.02 &11.38 $\pm$ 0.03 &12.28 $\pm$ 0.02 &1.31 $\pm$ 0.04 &0.41 $\pm$ 0.03 \\
2MASSW J1328550$+$211449 &L5 &14.20 $\pm$ 0.01 &13.01 $\pm$ 0.06 &13.84 $\pm$ 0.06 &1.20 $\pm$ 0.06 &0.37 $\pm$ 0.06 \\
2MASSW J1553214$+$210907 &L5.5 &14.67 $\pm$ 0.02 &13.26 $\pm$ 0.07 &13.93 $\pm$ 0.06 &1.41 $\pm$ 0.07 &0.74 $\pm$ 0.06 \\
2MASSI J0756252$+$124456 &L6 &14.89 $\pm$ 0.02 &13.33 $\pm$ 0.07 &14.56 $\pm$ 0.05 &1.55 $\pm$ 0.07 &0.32 $\pm$ 0.06 \\
2MASSW J0829570$+$265510 &L6.5 &14.87 $\pm$ 0.02 &13.23 $\pm$ 0.07 &14.18 $\pm$ 0.05 &1.64 $\pm$ 0.07 &0.68 $\pm$ 0.06 \\
2MASSI J1526140$+$204341 &L7 &13.91 $\pm$ 0.01 &12.59 $\pm$ 0.06 &13.50 $\pm$ 0.05 &1.32 $\pm$ 0.06 &0.41 $\pm$ 0.05 \\
2MASSI J0825196$+$211552 &L7.5 &13.02 $\pm$ 0.01 &11.34 $\pm$ 0.04 &12.43 $\pm$ 0.05 &1.68 $\pm$ 0.04 &0.59 $\pm$ 0.05 \\
2MASSW J1632291$+$190441 &L8 &14.01 $\pm$ 0.01 &12.58 $\pm$ 0.08 &13.52 $\pm$ 0.05 &1.43 $\pm$ 0.08 &0.49 $\pm$ 0.05 \\
SDSS 0837172$-$000018 &T1 &15.97 $\pm$ 0.04 &14.42 $\pm$ 0.11 &15.71 $\pm$ 0.09 &1.54 $\pm$ 0.12 &0.26 $\pm$ 0.10 \\
SDSS 1254539$-$012247 &T2 &13.95 $\pm$ 0.01 &12.12 $\pm$ 0.05 &13.62 $\pm$ 0.04 &1.83 $\pm$ 0.05 &0.34 $\pm$ 0.05 \\
SDSS 1021097$-$030420\tablenotemark{c} &T3 &15.40 $\pm$ 0.03 &13.53 $\pm$ 0.10 &15.53 $\pm$ 0.08 &1.86 $\pm$ 0.10 &$-$0.13 $\pm$ 0.08 \\
SDSS 1346465$-$003150\tablenotemark{c} &T6 &15.89 $\pm$ 0.03 &13.55 $\pm$ 0.06 &15.56 $\pm$ 0.07 &2.35 $\pm$ 0.07 &0.34 $\pm$ 0.08 \\
SDSS 1624144$+$002916 &T6 &15.66 $\pm$ 0.02 &13.22 $\pm$ 0.06 &15.43 $\pm$ 0.09 &2.44 $\pm$ 0.07 &0.24 $\pm$ 0.10 \\
\enddata

\tablenotetext{a}{Spectral Class:  M and L from \citet{kir99}; T from A. Burgasser (2001, private communication)}
\tablenotetext{b}{DENIS J1441-0945 is a binary recently resolved by Eduardo Mart\'\i n (2001, private communication).  This object is elongated in our images, and we estimate an approximate spatial separation of 0.42\arcsec  }
\tablenotetext{c}{Adjacent background galaxy removed prior to flux calculation}
\end{deluxetable}

\clearpage

\figcaption[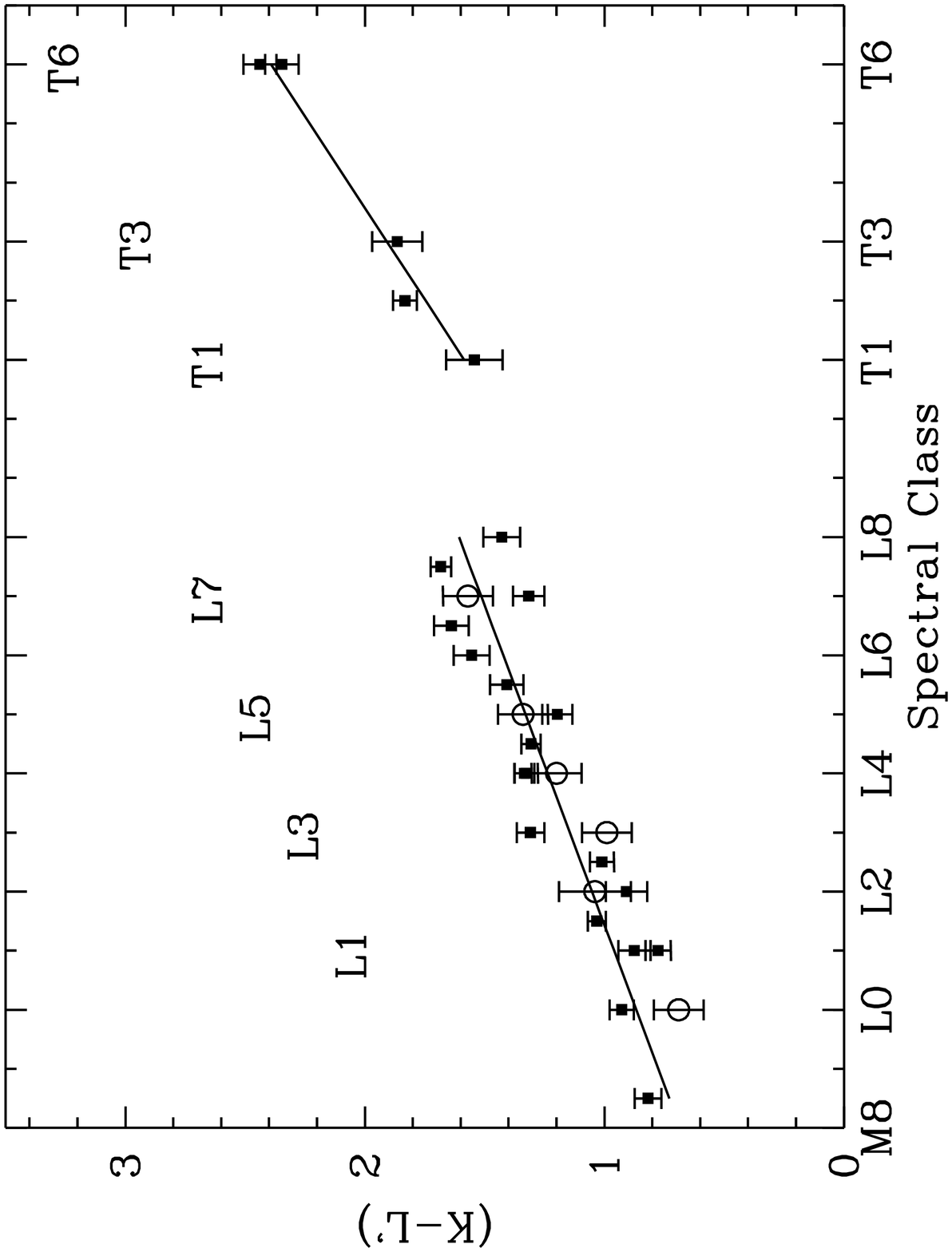]{({\it K-L$^\prime$}) colors as a function of spectral class.  The squares represent data from 
Table 1, and the six open circles are 2MASP 0345432+254023, Kelu-I, DENIS-P J1058-1548, GD 165B, DENIS-P J1228-1547AB, and DENIS-P
J0205-1159AB taken from \citet{leg01}. \label{fig1}}

\figcaption[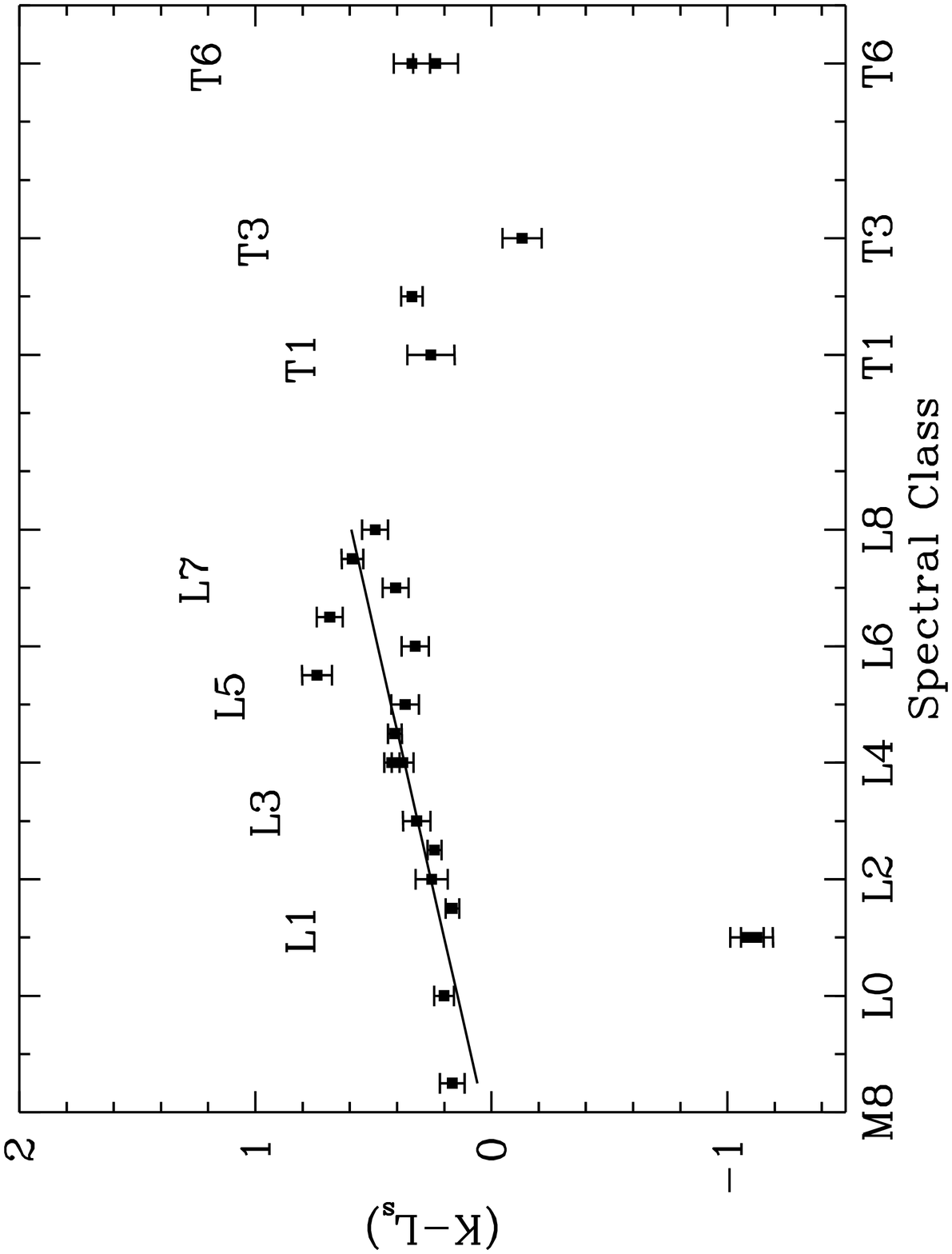]{({\it K}-$\rm {\it L_s}$) colors of Table 1 as a function of spectral class.  \label{fig2}}

\figcaption[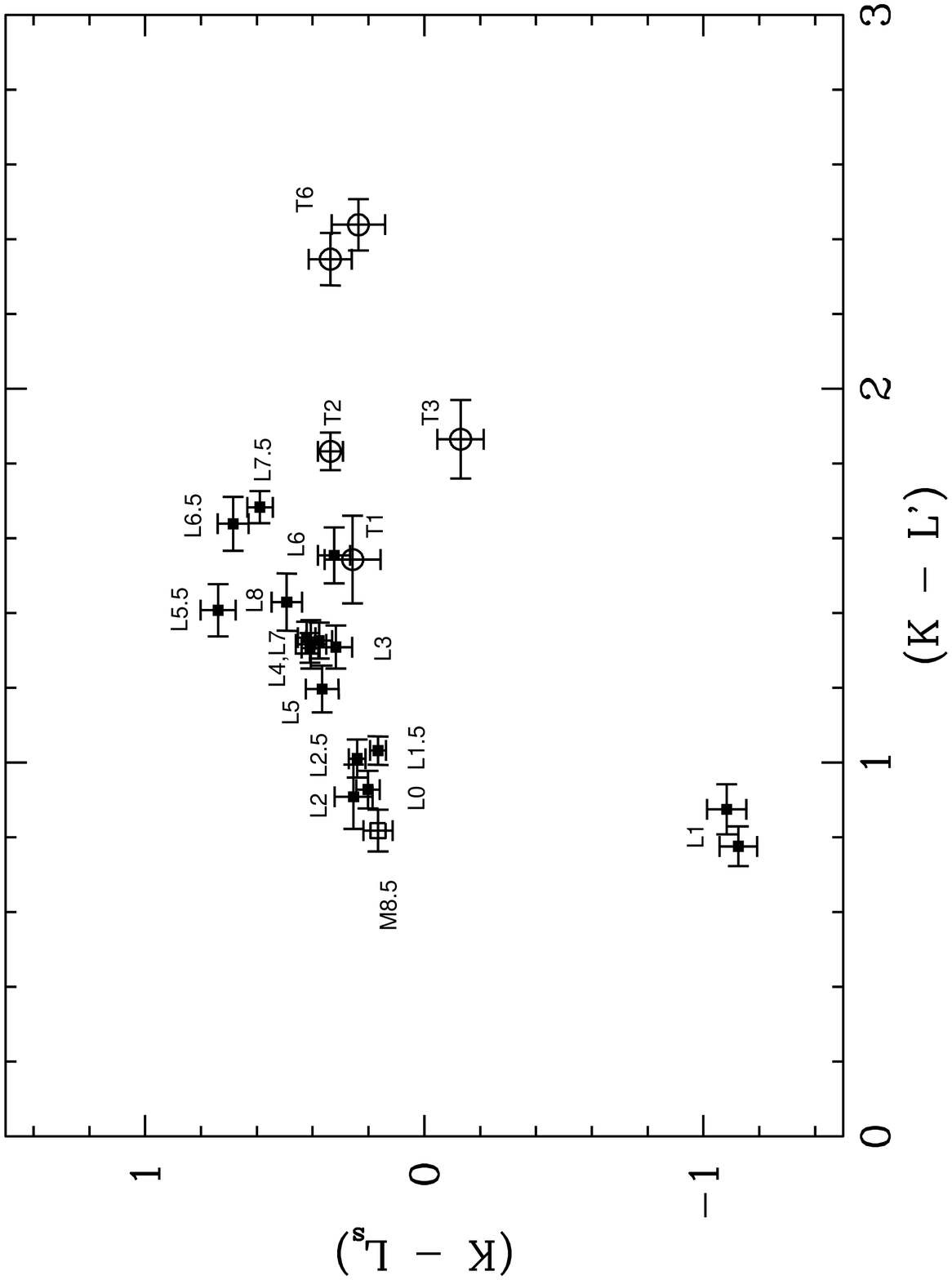]{({\it K}-$\rm {\it L_s}$) vs. ({\it K-L$^\prime$}) color-color diagram.  T dwarfs are denoted by open circles, L dwarfs
by filled squares and the lone M dwarf by an open square. \label{fig3}}

\end{document}